\documentstyle[12pt]{article}
\begin{document}
\renewcommand{\theequation}{\thesection.\arabic{equation}}

\vspace*{2cm}
\begin{center}
{\LARGE\bf Reggeon interactions in perturbative QCD   }
\end{center}
\vspace{2cm}
\begin{center}
{\large\bf R. Kirschner}\\{\large\it DESY - Institut f\"ur
Hochenergiephysik Zeuthen }\\{\large\it Platanenallee 6,
D-15735 Zeuthen, Germany }
\end{center}
\vspace*{5.0cm}
\begin{center}
{\bf Abstract }
\end{center}
We study the pairwise interaction of reggeized gluons and quarks in the
Regge limit of perturbative QCD. The interactions are represented as
integral kernels in the transverse momentum space and as operators in
the impact parameter space. We observe conformal symmetry and
holomorphic factorization in all cases.

\newpage
\section{Introduction}

The data on deep-inelastic scattering at small $x$ show the features
expected from the perturbative Regge asymptotics in QCD \cite{HERA}.
This provides the motivation to study the perturbative Regge region of
QCD in more detail and to apply more effort to go essentially beyond the
leading logarithmic approximation.

It is known that in the leading logarithmic approximation of
perturbative QCD the Regge asymptotics with vacuum quantum number
exchange is determined by the exchange of two reggeized gluons
\cite{BFKL}. In the case of meson quantum number exchange we have
instead
the exchange of reggeized quark and anti-quark (of different flavours)
\cite{RK86} \cite{RK94}. It is also known that the leading logarithmic
contribution does not give the true asymptotics. Unitarity corrections
become important with increasing energy. A minimal set of unitarity
corrections consists of the contributions from the exchange of an
arbitrary number of reggeized gluons with the same overall t-channel
quantum numbers. There the trajectories of the reggeized gluons and
quarks are to be taken in the leading logarithmic approximation. The
exchanged reggeons interact only pairwise and the effective vertices,
which determine this interaction, are just the same as in the
two-reggeon exchange of the leading logarithmic approximation.
This minimal improvement obeying unitarity in all sub-energy channels
can be called generalized leading logarithmic approximation.

Beyond the leading logarithmic approximation there are other
contributions not improving s-channel unitarity. They will not be
considered here. In the Regge asymptotics they are to be treated as
corrections to the reggeon interactions and to the reggeon-particle
vertices \cite{FL}.

In the case where the number of exchanged gluons and quarks does not
change, the equations for the partial waves, the reggeon Green function,
can written as a straightforward generalization of the two-reggeon case
Fig. 1 \cite{KwP} \cite{JB} \cite{L89}.
\begin{eqnarray}
\omega \ \ f_{\{ i_{\ell}\} }  = f^{(0)}_{\{i_{\ell} \} } +
\sum_{ i<j} {\cal H}^{(i,j)}_{\{i_{\ell} \} , \{j_{\ell} \} }
  f_{\{i^{\prime }_{\ell} \} }
\end{eqnarray}

The set $\{ i_{\ell} \} = (i_1,...,i_r ) $ labels the reggeons, where
$i = G $ stands for a reggeized gluon and $i = F $ or $ i = \overline F
$ stands for reggeized fermions in dependence on the helicity.
Corresponding to each reggeon $i$ the partial wave $ f $ carries an
index $\alpha_i $, labelling the state in the corresponding gauge group
representation and the argument $\kappa _i $, the transverse momentum,
or $ x_i $, the impact parameter of the reggeon. The term with $i$ and
$j$ in the sum in (1.1) is the contribution from the interaction between
the corresponding two reggeons.

The $r$-reggeon contribution to the partial-wave of the scattering
amplitude is obtained from the $r$-reggeon Green function by
convolution
with impact factors $\Phi $, depending on the scattering particles. The
angular momentum $j$ is related to $\omega $ by
\begin{equation}
j= 1 + \omega - { r_{F}\over 2}.
\end{equation}
$r_F $ is the number of exchanged fermions.

The present paper is devoted to the analysis of all two-reggeon
interactions ${\cal H }^{(i,j)} $ in QCD. The cases $i=j=G$ and $i=F, j=
\overline F $ are known from the studies of the two-reggeon exchange
with vacuum \cite{BFKL} \cite{L86} and meson quantum numbers
\cite{RK94}.
Besides of these there is the case $i=j= F$ of two exchanged fermions
carrying the same helicity and the case $i= F, j = G $ of an exchanged
fermion interacting with an exchanged gluon. The latter case appears
with two contributions, corresponding to the interaction via $s$-channel
gluon or fermion.

The graphical rules of the multi-Regge effective action \cite {KLS}
provide the framework for a simple derivation of ${\cal H}^{(i,j)} $. We
show that in all cases the eigenvalues have a form compatible with
conformal symmetry and holomorphic factorization. We consider the
representation both in transverse momentum and in impact parameter
spaces. In the first case ${\cal H}^{(i,j)} $ are written as integral
kernels and in the second case as operators. The operator representation
allows to write the interaction explicitely in a conform-symmetric and
factorizable form.

The ideas of conformal symmetry in the Regge asymptotics and the
operator approach appeared first in the study of gluon exchange
\cite{L86} \cite{L90} \cite{L93} and have been developed in application
to the fermion exchange ($i = F, j= \overline F$) \cite{RK94}. Here we
show that the methods allow an unified treatment of all reggeon
interactions.

\section{The interaction kernels}
\setcounter{equation}{0}

Consider the equation for the reggeon Green function $f$ (1.1)
represented graphically in Fig. 1.  The arguments referring to the
reggeons $\overline i_l $ are not affected by the operator in the
equation and will be suppressed. In the transverse momentum
representation we have
\begin{eqnarray}
{\cal H}^{(i,j)} f_{i_1...i...j...i_r} =
\frac{g^2}{(2 \pi )^3} \int d\kappa^{\prime}_i d\kappa^{\prime }_j
         \delta (\kappa_i + \kappa_j - \kappa^{\prime }_i -
\kappa^{\prime }_j ) \cr
\left[(T_i \otimes T_j)_{\cal H} \ \
{\cal H}_{i,j}
f_{i_1...i...j...i_r } +    (T_i \otimes T_j )_{\cal G}  \ \
 {\cal G}_{i,j} f_{i_1 ... j...i...i_r } \right]
\end{eqnarray}
  $T^{a}_i , a= 1,...,N,$ are the
generators of the $SU(N) $ representation of the reggeon $i$.
The first contribution in (2.1) corresponds to the interaction via an
s-channel gluon. Therefore the gauge group matrix is obtained from the
generators by summing over the gluon colour states $a$.
\begin{equation}
(T_i \otimes T_j )_{{\cal H} \{\alpha_{\ell} , \alpha^{\prime }_{\ell}
\} } = \prod_{\ell \not= i,j}^{r}
\delta_{\alpha_{\ell} , \alpha_{\ell}^{\prime  }} \cdot
(T^{a} )_{\alpha_i,\alpha_i^{\prime }}
(T^{a} )_{\alpha_j,\alpha_j^{\prime }}
\end{equation}
The second contribution in (2.1) corresponds to the interaction via an
s-channel fermion. In this case the gauge group matrix os obtained by
summing over the fermion colour state $\alpha $, e.g. if $i=G, j=F $,
\begin{equation}
(T_i \otimes T_j )_{{\cal G} \{\alpha_{\ell} ,
\alpha^{\prime }_{\ell} \} } = \prod_{\ell \not= i,j}^{r}
\delta_{\alpha_{\ell} , \alpha_{\ell}^{\prime  }} \cdot
(T^{\alpha_i^{\prime }} )_{\alpha_i,\alpha }
(T^{\alpha_j} )_{\alpha,\alpha_j^{\prime }}
\end{equation}
The overall gauge group state in the $t$ channel is the singlet state.
(1.1) is valid only in this case.

There are important special cases, when the interaction in gauge-group
representation space reduces to multiplication by a number. This
simplification happens in the two- and three-gluon exchange and becomes
an approximation at large N.

The kernels are functions of $\kappa_i, \kappa_j, \kappa^{\prime  }_i,
\kappa^{\prime }_j $ and as arguments of $f$ we have $\kappa^{\prime },
\kappa^{\prime }_j $ and $ \kappa_{\ell} $ for $ \ell \not= i,j $. In
the momentum representation we prefer to modify the definition of the
reggeon Green function $f$ in such a way that it includes from the
propagator of the $r$-reggeon state,
\begin{equation}
\left[ \omega - \sum_{\ell = 1}^{r} \alpha_{\ell} (\kappa_{\ell} )
\right]^{-1} \prod_{1}^{r} d_{\ell}(\kappa_{\ell}) ,
\end{equation}
only the first factor but not the $t$-channel propagators $ d_{\ell}
(\kappa_{\ell} ) , \ell = 1,...,r $,
\begin{equation}
d_{G} (\kappa) = \vert \kappa \vert^{-2}, \ \ \ d_{F} (\kappa ) =
\kappa^{* -1 }, d_{\overline F} = \kappa^{-1}.
\end{equation}
We use the complex notation for the two-dimensional transverse momenta,
$\kappa = \kappa_1 + i \kappa_2 $.
$\alpha_i (\kappa ) $ determine the trajectories of the reggeized gluons
($i = G$) or quarks ($ i = F, \overline F $). The gluon trajectory is
\begin{eqnarray}
1 +  N \alpha _{G} (\kappa ),  \ \ \ \ \ \
\alpha _{G} (\kappa ) = \frac{g^2}{2(2 \pi )^3}
\int {d^2\kappa^{\prime } \vert \kappa \vert^2  \over
\vert \kappa - \kappa^{\prime } \vert^2 \vert \kappa^{\prime } \vert^2 }
\end{eqnarray}
and the fermion trajectories are
\begin{eqnarray}
&\frac{1}{2} + C_2  \alpha_F (\kappa ) ,  \ \ \ \ \ \ \
&\alpha_F (\kappa ) = \frac{g^2}{(2 \pi )^3}
\int {d^2\kappa^{\prime }  \kappa^{*}  \over
\vert \kappa - \kappa^{\prime } \vert^2  \kappa^{\prime * }  } ,
\cr
&\frac{1}{2} + \alpha_{\overline F} (\kappa ), \ \ \ \ \  \
&\alpha_{\overline F} (\kappa ) = (\alpha_F (\kappa ) )^{*}.
\end{eqnarray}
The interaction kernels can be derived easily by applying the effective
graphical rules \cite{KLS}. We use the effective vertices (Fig. 2a)
$V_{ij}^{\lambda }$, where $ \lambda $ denotes the helicity of the
produced gluon,
\begin{eqnarray}
&V_{GG}^{+} = \kappa \kappa^{\prime *}, \ \ \ \ \ \ &V_{GG}^{-} =
\kappa^{*} \kappa^{\prime },
\cr
&V_{FF}^{-} = \kappa^{\prime *} , \ \ \ \ \ \ \ &V_{FF}^{+} =
\kappa^{*}, \cr
&V_{FG} = \kappa^{\prime *}, \ \ \ \ \ \ \ \ &V_{GF} = \kappa^{*}
\end{eqnarray}
By complex conjugation one obtains the corresponding vertices for the
opposite helicities of the fermions and of the produced gluon.

We need also the transverse momentum factors $D_{ij} (\kappa -
\kappa^{\prime } ) $ in the propagators of the s-channel gluon or quark.
For the gluon we have
\begin{equation}
D_{GG} = D_{FF} = D_{\overline F \overline F} = \vert \kappa -
\kappa^{\prime } \vert^{-2}
\end{equation}
and for fermions
\begin{equation}
D_{FG} = - D_{GF} = (\kappa - \kappa^{\prime } )^{* -1}, \ \ \ \
D_{\overline F G} = - D_{G \overline F} = (\kappa - \kappa^{\prime
})^{-1}.
\end{equation}
Working with partial waves the longitudinal momentum integral of the
loop in the last grapg in Fig.1 can be calculated.  We are left
with the transverse momentum intergal of the form (2.1) with the kernel
obtained from the graph Fig.2b with the vertices (2.8)  and the
propagators (2.5) and (2.9), (2.10).
\begin{eqnarray}
&{\cal H}^{(0)}_{ij,i^{\prime }j^{\prime } }(\kappa_i, \kappa_j;
\kappa_{i^{\prime }}, \kappa_{j^{\prime }} ) =  \cr
&\sum_{\lambda = +,-}
V_{i i^{\prime }}^{-\lambda }(\kappa_i, \kappa_{i^{\prime }} )
D_{i i^{\prime }}(\kappa_i - \kappa_{i^{\prime }} )
V_{j j^{\prime }}^{\lambda }(\kappa_j, \kappa_{j^{\prime }} )
d_{i^{\prime }} (\kappa_{i^{\prime }})
d_{j^{\prime }} (\kappa_{j^{\prime }})
+ (i \leftrightarrow j )
\cr
&{\cal H}^{(0)}_{ij,i^{\prime } j^{\prime }} =
\delta_{i i^{\prime }} \delta_{j j^{\prime }} {\cal H}^{(0)}  +
\delta_{i j^{\prime }} \delta_{j i^{\prime }} {\cal G}^{(0)}
\end{eqnarray}
In the case of interaction via s-channel gluon there is a sum over its
helicity states.

In the case $i = i^{\prime } = F, j = j^{\prime } = \overline F  $ the
double logarithmic divergencies require a more careful treatment of the
longitudinal momentum integration, which leads to additional
$\omega$-dependent factors in the final expression. We shall not repeat
the details given in \cite{RK94} and quote just the result.

The sum over all loop contributions enters the equation Fig. 1
originally multiplied by the angular momentum factor of the $r$-reggeon
propagator (2.4). We multiply both sides by the inverse of this factor,
$ [\omega - \sum \alpha_{\ell} ] $, and add $\sum \alpha_{\ell} f $.
Then
the latter contribution can be included into the kernel ${\cal H}_{ij,
i^{\prime }j^{\prime }}$
(2.11) with the same gauge group operator in front (2.1). This is a
non-trivial
step, which can be done only because the overall gauge group state in
the t-channel is the singlet one.

The resulting kernels as they enter the equation (1.1) in the transverse
momentum representation (2.1) are obtained in the following form. We
replace $\kappa_{i} \rightarrow \kappa_1,
\kappa_{i^{\prime }} \rightarrow \kappa_1^{\prime },
\kappa_{j} \rightarrow \kappa_2,
\kappa_{j^{\prime }} \rightarrow \kappa_2 $.
\begin{eqnarray}
{\cal H}_{GG} = \vert \kappa_1 - \kappa_1^{\prime } \vert^{-2 }
\left ({\kappa_1 \kappa_2^{*} \over \kappa_1^{\prime}
\kappa_2^{\prime *}} +
{\kappa_1^{*} \kappa_2 \over \kappa_1^{\prime *} \kappa_2^{\prime } }
\right )
-  (\alpha_G(\kappa_1 ) +  \alpha_G (\kappa_2 ) )
\delta (\kappa_1 -\kappa_1^{\prime}  ), \cr
{\cal H}_{F \overline F}^{(\omega )} = \vert \kappa_1 - \kappa_1^{\prime
} \vert^{-2}
\left ( \left \vert { \kappa_1 \over \kappa_1^{\prime } } \right
\vert^{\omega } +
 \left \vert { \kappa_1^{\prime } \over \kappa_1 } \right \vert^{\omega}
{\kappa_1^{*} \kappa_2 \over \kappa_1^{\prime *} \kappa_2^{\prime } }
\right )
-  (\alpha_F(\kappa_1 ) +  \alpha_F^{*} (\kappa_2 ) )
\delta (\kappa_1 -\kappa_1^{\prime } ), \cr
{\cal H}_{FG} = \vert \kappa_1 - \kappa_1^{\prime } \vert^{-2}
\left (
{\kappa_1^{*} \kappa_2 \over \kappa_1^{\prime *} \kappa_2^{\prime } }  +
{\kappa_2^{*} \over \kappa_2^{\prime *} } \right )
- (\alpha_F(\kappa_1 ) +   \alpha_G (\kappa_2 ) )
\delta (\kappa_1 -\kappa_1^{\prime } ), \cr
{\cal H}_{FF} = \vert \kappa_1 - \kappa_1^{\prime } \vert^{-2}
\left (
{\kappa_2^{*}  \over \kappa_2^{\prime *}  }  +
{\kappa_1^{*} \over \kappa_1^{\prime *} } \right )
- (\alpha_F(\kappa_1 ) +   \alpha_F (\kappa_2 ) )
\delta (\kappa_1 -\kappa_1^{\prime } ), \cr
{\cal G}_{FG} = \vert \kappa_1 - \kappa_1^{\prime } \vert^{-2}
\left ({\kappa_1 \over \kappa_1^{\prime } } - 1 \right ).
\end{eqnarray}
The cases with the replacement $ F \leftrightarrow \overline F $
are obtained by complex conjugation.

\section{The eigenvalues of the interaction kernels}
\setcounter{equation}{0}

We study the special case if the momentum transfer in the
$t$-subshannel (i,j) vanishes, $\kappa_i = - \kappa_j = \kappa,
\kappa_i^{\prime } = - \kappa_j^{\prime } = \kappa^{\prime } $. The
second term in (2.1) can be understood as $ P_{ij} {\cal G}_{ij} $,
where $ P_{ij} $ is the operator of the permutation of the reggeon $i$
and $j$. Therefore it is reasonable to look at the eigenvalue problem
for ${\cal G}_{ij}$ as well as for ${\cal H}_{ij} $.
\begin{eqnarray}
\int d^2 \kappa^{\prime } {\cal H}_{ij} (\kappa , \kappa^{\prime } )
f (\kappa^{\prime} ) = \pi  \Omega_{ij} \ \  f(\kappa ), \cr
\int d^2 \kappa^{\prime } {\cal G}_{ij} (\kappa , \kappa^{\prime } )
f (\kappa^{\prime} ) = \pi  \Omega_{i/j} \ \  f(\kappa ).
\end{eqnarray}
In the subchannel with vanishing momentum transfer we have rotation
symmetry in the plane of transverse momenta. The appropriate complete
orthogonal set of functions is
\begin{equation}
\phi_{n, \nu } (\kappa )  = \vert \kappa \vert ^{-1 + 2 i \nu }
\left ({\kappa^* \over \vert \kappa \vert } \right )^n  ,
\end{equation}
parametrized by $\nu $, running over the real axis, and by $n$, taking
all integer or all half-integer values in dependence of whether the
fermion number in the sub-channel $(ij) $ is even or odd.

We choose the eigenfunctions $f^{n, \nu }$  in such a way that their
orthogonality relation is written with the propagators of the particles
$i$ and $j$ as a weight. Therefore we define
\begin{equation}
f^{n, \nu } = \left ( d_i (\kappa ) d_j (\kappa ) \right )^{-1/2 }
\phi_{n,\nu }.
\end{equation}
In the calculation it is convenient to extend the dimension to $2 +
2\epsilon $. The pole terms in $\epsilon $ cancel, since the kernels are
infrared finite, and we obtain
\begin{eqnarray}
&\Omega_{F\overline F} = 4 \psi (1)   \cr
&- \frac{1}{2} \left (
\psi (m +\frac{1 - \omega}{2}) +\psi (1- m + \frac{1 - \omega}{2})
+ \psi (m - \frac{1 - \omega}{2}) +\psi (1- m -\frac{1 - \omega}{2})
\right ) \cr
&- \frac{1}{2} \left (
 \psi (\tilde m + \frac{1 - \omega}{2}) +\psi (1- \tilde m + \frac{1 -
\omega}{2} )
 + \psi (\tilde m - \frac{1 - \omega}{2}) +\psi (1- \tilde m - \frac{1 -
\omega}{2}) \right ) , \cr
&\Omega_{FG} = 4 \psi (1) - \psi (m) - \psi (1-m)   \cr
&- \frac{1}{2} \left (
 \psi (\tilde m + {1  \over 2}) +\psi (1- \tilde m + {1  \over 2})
 + \psi (\tilde m - {1 \over 2}) +\psi (1- \tilde m - {1 \over 2})
 \right ) , \cr
&\Omega_{GG} = 4 \psi (1)
-\psi (m) -\psi (1- m) - \psi (\tilde m) - \psi ( 1- \tilde m ), \cr
&\Omega_{FF } = \Omega_{GG}, \ \ \ \ \ \ \ \ \
\Omega_{F/G} = (\tilde m - {1 \over 2 } )^{-1}.
\end{eqnarray}
We have used the digamma function,
\begin{eqnarray}
\psi (z) = {d \over dz} \ln \Gamma (z) =
\psi (1) -\sum_{\ell = 0}^{\infty} ({1 \over \ell + z } - {1 \over \ell
+1 }).
\end{eqnarray}
The notations
\begin{equation}
m = \frac{1}{2} + i \nu + {n \over 2},
\tilde m = \frac{1}{2} + i \nu - {n \over 2}
\end{equation}
appear in the following as conformal weights. All eigenvalue functions
(3.4) can be written as sums of two functions, one depending on the
eigenvalue $m(1-m)$ of the holomorphic and the other depending on the
eigenvalue $\tilde m (1- \tilde m) $ of the anti-holomorphic Casimir
operator of the linear conformal transformations. This is the necessary
condition for conformal symmetry and holomorphic factorization of the
equation.

We define
\begin{equation}
\chi_{\Delta } (z) = \sum_{\ell = 0}^{\infty}
\left ( {2 (\ell + \Delta )+1 \over (\ell + \Delta )(\ell + \Delta +1)
+ z }
- {2 \over \ell + 1 } \right )
\end{equation}
and observe from (3.5)
\begin{equation}
\chi_{\Delta} (m(1-m)) = 2 \psi (1) - \psi (m + \Delta )
- \psi ( 1 - m + \Delta )
\end{equation}
to obtain the desired result for the eigenvalue functions,
\begin{eqnarray}
&\Omega_{F\overline F}  = \frac{1}{2} \left (
\chi_{{1-\omega \over 2}} ( m(1-m) )  +
\chi_{{\omega -1 \over 2}} ( m(1-m) )  +
\chi_{{1-\omega \over 2}} (\tilde m (1- \tilde m ) +
\chi_{{\omega - 1 \over 2}} (\tilde m (1- \tilde m )
\right ), \cr
&\Omega_{GG} = \chi_{0} ( m(1-m) ) + \chi_{0} (\tilde m (1- \tilde m )
),                     \cr
&\Omega_{FG} = \chi_{0} ( m(1-m) )  + \frac{1}{2} \left (
\chi_{{1 \over 2}} (\tilde m (1- \tilde m )   +
\chi_{-{ 1 \over 2}} (\tilde m (1- \tilde m )   \right ), \cr
&\Omega_{GG} = \chi_{0} ( m(1-m) ) + \chi_{0} (\tilde m (1- \tilde m )
),                     \cr
&\Omega_{F/G} = \left ( - \tilde m (1 - \tilde m ) +\frac{1}{4}
\right )^{-1/2 }.
\end{eqnarray}
We find that $\Omega_{FG} $ consists of the holomorphic part of
$\Omega_{GG} $ and the anti-holomorphic part of $\Omega_{F \overline
F}$,
where in the latter $\omega$ is put to zero. $\Omega_{FF} $ coincides
with $\Omega_{GG}$. The holomorphic part of $\Omega_{F/G} $ vanishes.

\section{Operators in the impact parameter space}
\setcounter{equation}{0}

We study the reggeon Green function in the impact parameter
representation, $f(\omega; x_1,...,x_r)$, keeping the symbol $f$ also
for the Fourier transformed function. We use an operator representation
for the equation (1.1).
\begin{equation}
{\cal H}^{(ij)} f_{i_1...i...j...i_r} = \frac{g^2}{8 \pi^2 }
\left ( (T_i\otimes T_j)_{\cal H} \ \
\hat {\cal H}_{ij} f_{i_1...i...j...i_r}
+  (T_i\otimes T_j)_{\cal G}   \ \
  \hat {\cal G}_{ij} f_{i_1...j...i...i_r} \right )
\end{equation}
The operatores $\hat {\cal H}_{ij}$ and $\hat {\cal G}_{ij} $ are
functions of the elementary operators of multiplication with $x_i$ and
$x_j$ and differentiations $\partial_i$ and $\partial_j $. They can be
read off from the kernels (2.12) by the following simple substitution
rules. The momenta are replaced by derivatives,
$\kappa_1 \rightarrow \partial_1^*,
\kappa_1^* \rightarrow \partial_1, $
and $\delta (\kappa_1 - \kappa_2 )$ by 1.
The propagators of the $s$-channel particles are replaced as
\begin{eqnarray}
D_G (\kappa_1 - \kappa_1^{\prime }) =
{1 \over \vert \kappa_1 - \kappa_1^{\prime } \vert^2 }
\rightarrow  -\ln \vert x_{12}^2 \vert + \psi(1), \cr
D_F (\kappa_1 - \kappa_1^{\prime }) =
{1 \over ( \kappa_1 - \kappa_1^{\prime } )^* } \ \ \ \ \
\rightarrow  ( x_{12}^* )^{-1},
\end{eqnarray}
where $x_{12} = x_1 - x_2 $. The trajectories are replaced as
\begin{equation}
\frac{1}{2} \alpha_G (\kappa_1) = \alpha_F (\kappa_1 )  \ \ \ \
\rightarrow  -\ln ( \partial_1 \partial_1^* ).
\end{equation}
(4.2) and (4.3) can be understood in dimensional regularization. The
poles in $\epsilon $ from the $s$-channel gluon propagator (4.2) and
from the trajectories cancel in the operators.

 The resulting operators decompose into sums of holomorphic (acting only
on $x_1, x_2$ ) and anti-holomorphic parts.
\begin{eqnarray}
\hat {\cal H}_{GG} = H_G + H_G^* , \cr
\hat {\cal H}^{(\omega )}_{F\overline F} = H^{(\omega )}_F + P_{12}
H^{(\omega ) *}_F  P_{12} , \cr
\hat {\cal H}_{FF} = H_G + \tilde H_F^* , \cr
\hat {\cal H}_{FG} = H_G + P_{12} H_F^{(0) *} P_{12} , \cr
\hat {\cal G}_{FG} = (x_{12}^* \partial_2^* )^{-1} = - P_{12} D_1^{*
-1} P_{12}.
\end{eqnarray}
$P_{12} $ denotes the operator of permutation of $x_1$ and $x_2$. We use
the notation $D_1 = x_{12} \partial_1$ and $D_2 = x_{12} \partial_2 $.

The operators $H_G$ and $H_F^{(\omega )}$ have been encountered before
\cite{L90} \cite{L93} \cite{RK94}.
\begin{eqnarray}
H_G = 2 \psi (1) -
\partial_1^{-1} \ \ \ln x_{12} \ \ \partial_1
- \partial_2^{-1} \ \ \ln x_{12} \ \ \partial_2
- \ln \partial_1 \ \  -\ln \partial_2 ,  \cr
H_F^{(\omega )} = 2 \psi (1) -
\partial_1^{-1 + \omega /2} \ \ \ln x_{12} \ \ \partial_1^{1- \omega /2}
- \ \ \ln \partial_1
- \partial_2^{-\omega /2} \ \ \ln x_{12} \ \ \partial_2^{\omega /2}
 \ \  -  \ \ \ln \partial_2 .
\end{eqnarray}
The new operators are $\tilde H_F^{*} $, being the complex conjugate of
\begin{equation}
\tilde H_F = 2 \psi (1) - 2 \ln x_{12} - \ln \partial_1 - \ln
\partial_2,
\end{equation}
and the simple operator ${\cal G}_{FG} $ given explicitely in (4.5).
We express $H_G, H_F$ and $\tilde H_F $ as $\psi$-functions of $D_1 =
x_{12} \partial_1 $ and $D_2 = x_{12} \partial_2 $. This representation
is obtained by observing that $x^2 \partial $ and $\partial $ can be
written as similarity transformations of $x$ and $x^{-1} $,
respectively.
\begin{eqnarray}
x^2 \partial = \Gamma (x \partial ) x (\Gamma (x \partial ) )^{-1} ,
\cr
\partial = (\Gamma (x \partial + 1))^{-1}  x^{-1} \Gamma (x \partial +
1).
\end{eqnarray}
The operators as functions of $x \partial $ are determined actually only
up to periodic functions with the period 1 of the same argument. This
ambiguity matters if we apply (4.7) to logarithms. The relations imply
the identity
\begin{equation}
\ln (x^2 \partial ) - \ln x  =  \partial^{-1}  \ln x \partial  -
\ln \partial
\end{equation}
and lead to the following representation,
\begin{eqnarray}
&H_G = 2 \psi (1) -
\frac{1}{2} \left ( \psi (D_1 ) + \psi (1 - D_1 ) + \psi (-D_2) + \psi
(1 + D_2) \right ), \ \ \cr
&H_F^{(\omega )} = 2 \psi (1) -
\frac{1}{2} \left ( \psi (- D_1 + {\omega \over 2}) + \psi (1 - D_1
-{\omega \over 2}) \right ) +   \cr
& \frac{1}{2} \left ( \psi (-D_2 +
{\omega \over 2}) + \psi (1 + D_2 - {\omega \over 2}) \right ), \ \ \cr
&\tilde H_F = 2 \psi (1) -
\frac{1}{2} \left ( \psi (-D_1 ) + \psi ( D_2 ) + \psi (1 + D_1) +
\psi (1 - D_2) \right).
\end{eqnarray}

\section{Conformal symmetry}
\setcounter{equation}{0}

The result (3.9) about the eigenvalues tells us that the operators can
be written in a conform-symmetric form. We shall express the holomorphic
operators in terms of the holomorphic Casimir operator of the linear
conformal group.

 Starting with the form (4.5), (4.6) of the operators and applying the
relation (4.8) we observe the following behaviour under conformal
inversions ${\cal I}$.
\begin{eqnarray}
{\cal I} H_G {\cal I} = H_G,  \ \ \ \ \ \ \ \ \
{\cal I} H_F^{(0)} {\cal I} = x_2 H_F^{(0)} x_2^{-1},  \cr
{\cal I} \tilde H_F {\cal I} = x_1 x_2 \tilde H_G (x_1 x_2 )^{-1}, \ \ \
{\cal I} P_{12} D_1 {\cal I} = x_2 (P_{12} D_1 ) x_2^{-1}.
\end{eqnarray}
The equation with these operators is conform-symmetric if the Green
function $f$ transforms correspondingly,  i.e. as a correlator with
operators in the points $x_1, x_2 $ of the conformal weights $(\Delta_1,
\tilde \Delta_1 ), (\Delta_2, \tilde \Delta_2 )$.
\begin{equation}
\Delta_{\ell} = \frac{1}{2} (\delta_{\ell} + s_{\ell} ), \ \ \ \ \ \ \ \
\tilde \Delta_{\ell} = \frac{1}{2} (\delta_{\ell} - s_{\ell} ).
\end{equation}
$\delta_{\ell}, \ell =1,2 $ are the scaling dimensions and $s_{\ell}$
are the conformal spins. From (5.1) and (4.5) we see that the
conformal operator corresponding to a reggeized gluon has
\begin{equation}
\delta_G = s_G = 0
\end{equation}
and the conformal operators representing fermionic reggeon have in
dependence on the helicity have
\begin{equation}
\delta_F = - s_F = \frac{1}{2},
\delta_{\overline F} =  s_{\overline F} = \frac{1}{2}.
\end{equation}
The eigenvalue problem can also be studied in the operator
representation. Because of the conformal symmetry there is a simple
solution without restrictions on the momentum transfer.
The eigenfunctions are the conformal 3-point functions,
\begin{equation}
E_{\delta_1, s_1, \delta_2, s_2}^{(n,\nu )} =
\langle \phi_{\delta_1, s_1} (x_1) \phi_{\delta_2, s_2 } (x_2)
{\cal O}^{(n,\nu )} (x_0 ) \rangle ,
\end{equation}
which are determined up to a factor by the scaling dimensions $\delta_1,
\delta_2, \frac{1}{2} + i \nu $ and the conformal spins $s_1, s_2, n$
\cite{AP} \cite{L86}. For given $\delta_1, \delta_2, s_1, s_2$,
characterizing
the interacting reggeons ( compare (5.3), (5.4) ), they form a twofold
overcomplete basis of functions of two impact parameters $x_1,x_2$. The
set is parametrized by the scaling dimension $\nu$, running over the
real axis, the conformal spin $n$, taking all integer values for $s_1 +
s_2$ integer or all half-integer values for $s_1 + s_2 $ half-integer,
and by the position $x_0$, running over the impact parameter space.
The eigenvalues coincide with the above results (3.4), (3.9).

The explicite form of the 6 generators $M_{12}^{\pm}, M_{12}^{(0)}
\tilde M_{12}^{\pm}, \tilde M_{12}^{(0)} $ of the coformal
transformations acting on functions of $x_1$ and $x_2 $ depends on the
conformal weights $(\Delta_1, \tilde \Delta_1 ),
(\Delta_2, \tilde \Delta_2 ) $.
\begin{eqnarray}
M_{12}^{+} = x_1^2 \partial_1 + 2 \Delta_1 x_1  +
x_2^2 \partial_2 + 2 \Delta_2 x_2 , \cr
M_{12}^{-} =  \partial_1 + \partial_2, \cr
M_{12}^{(0)} = x_1 \partial_1 +  \Delta_1  + x_2 \partial_2 + \Delta_2.
\end{eqnarray}
The anti-holomorphic generators are obtained by complex conjugation and
by replacing   $\Delta_1, \Delta_2 $ by $\tilde \Delta_1, \tilde
\Delta_2 $.  The explicite form of the holomorphic Casimir operator
depends on the weights $\Delta_1, \Delta_2 $ as follows.
\begin{eqnarray}
C_{\Delta_1 \Delta_2} =
- M_{12}^{(0) 2}  + \frac{1}{2} (M_{12}^{+} M_{12}^{-} +
M_{12}^{-} M_{12}^{+} ) \cr
= x_{12}^2 \partial_1 \partial_2 + 2 x_{12} (\Delta_1 \partial_2 -
\Delta_2 \partial_1 ) +                         (\Delta_1 + \Delta_2 )
(1 - \Delta_1 -\Delta_2 ).
\end{eqnarray}
For values of $\Delta_1, \Delta_2 $ differing by $\frac{1}{2} $ the
Casimir operator can be expressed as a square of a simpler operator plus
a constant, a property related to supersymmetry. We need the cases
$\delta_1 = 0, \delta_2 = \frac{1}{2} $ and
$\delta_1 = \frac{1}{2}, \delta_2 = 0 $.
\begin{eqnarray}
C_{0 \frac{1}{2} } = - A_{F2}^2  + \frac{1}{4},  \ \ \ \
A_{F2} = P_{12} x_{12} \partial_1, \cr
C_{\frac{1}{2} 0} = - A_{F1}^2  + \frac{1}{4},   \ \ \ \
A_{F1} = P_{12} x_{12} \partial_2 .
\end{eqnarray}
We find immediately from (4.5) that the interaction by fermion exchange
is expressed in terms of these operators,
\begin{equation}
P_{12} {\cal G}_{FG} = (A_{F1}^{*} )^{-1} ,
\end{equation}
in agreement with the eigenvalues (3.4), (3.9).

We look for the expressions of the remaining operators in terms of the
Casimir operators starting from (4.9). We are aware of the ambiguity in
this representation because of the non-uniqueness of the similarity
transformation (4.7). The tranformations, which we are going to
perform,
are valid in general only approximately on a restricted class of
functions. In the $\rho q $ representation these functions should be
small outside the region $\rho q \ll 1$. The $\rho q $ representation is
obtained from the impact parameter representation $f (x_1,x_2) $ by
identifying $\rho $ with $x_{12} = x_1 - x_2$ and by Fourier
transformation with repect to $R = (x_{10} + x_{20})/2 $. The Fourier
conjugate variable to $R$ is the momentum transfer $q$. Acting on this
class of functions we have for the operator
\begin{equation}
D_1 + D_2 \ll 1.
\end{equation}
Within this approximation we obtain from (4.9) the expression in terms
of the Casimir operators using (3.5). Despite of the approximation the
result is valid in general, without restictions on the functions,
because of the conformal symmetry.
\begin{eqnarray}
&H_G = \chi_0 (C_{0 0} ) , \cr
&H_F^{(\omega)} = \frac{1}{2} (
\chi_{{1 -\omega \over 2}} (C_{0 \frac{1}{2} })  +
\chi_{{\omega - 1 \over 2}} (C_{0 \frac{1}{2} })  ), \cr
&\tilde H_F = \chi_{0} ( C_{\frac{1}{2} \frac{1}{2} } ).
\end{eqnarray}
These results confirm the results (3.9) about the eigenvalues.

\section{Discussions}
\setcounter{equation}{0}

The Regge asymptotics in perturbative QCD can be represented in terms of
reggeized gluons and quarks in the exchange channel which interact by
emitting and absorbing s-channel gluons and quarks. This representation
is quite useful although those reggeons have no meaning independent of a
regularization  because their trajectories are infrared divergent.
However in gauge group singlet channels these divergencies cancel
against divergencies in the reggeon interactions. Therefore in these
channels the reggeon interaction can be represented in terms of infrared
finite operators.

In the generalized leading logarithmic approximation we impose the
conditions of multi-Regge kinematics on all s-channel intermediate
states. In this case the reggeon interact only pairwise.

We have studied all cases of these two-reggeon interactions that occur
in QCD. We have represented them as intergral kernels in transverse
momenta and as operators in the impact parameter space. The eigenvalues
can be obtained in the momentum representation in the limit of vanishing
momentum transfer. The form of the resulting eigenvalue functions is
compatible with the properties of holomorphic factorization and
conformal symmetry. The operator approach in the impact parameter space
allows to represent the two-reggeon interaction operators as sums of
holomorphic and anti-holomorphic parts. Moreover these parts are
functions of the Casimir operator of the holomorphic or anti-holomorphic
linear conformal transformations, respectively. More details about this
approach will be published elsewhere \cite{KL94}.

It is remarkable that the treatment invented for the gluon exchange
extends to the case involving fermions. Whereas the reggeized gluons are
represented by operators with vanishing conformal dimension and spin,
the reggeized fermions correspond to conformal dimension $\frac{1}{2} $
and spin $\pm \frac{1}{2} $.
This result provides a further piece of information about the simplicity
and the symmetry of perturbative QCD in the Regge limit.

\vspace*{2cm}

$\mbox{ }$ \\
{\Large\bf Acknowledgements} \\
$\mbox{ }$ \\
The author is grateful to L.N. Lipatov and L. Szymanowski for
discussions.

\vspace{2cm}

\noindent{\Large\bf Figure captions}
\vspace{1cm}

\begin{tabular}{ll}

Fig. 1 & The equation for the r-reggeon Green function.  \\
       & The horizontal lines represent reggeized gluons or quarks.  \\
       & The vertical line is a gluon (${\cal H}_{ij}$) or a quark
(${\cal G}_{FG}$).\\
       & In the last term a sum over $i,j$ is understood. \\

Fig. 2 & Graphical rules for the interaction kernels. \\
       & a) Effective vertex.  \\
       & b) Two-reggeon interaction. \\
\end{tabular}

\newpage

\input FEYNMAN


\begin{picture}(50000,21000)
\drawline\fermion[\E\REG](3000,5000)[5000]
\drawline\fermion[\N\REG](\particlebackx,\particlebacky)[10000]
\drawline\fermion[\W\REG](\particlebackx,\particlebacky)[5000]
\drawline\fermion[\S\REG](\particlebackx,\particlebacky)[10000]
\put(5000,10000){$ f $}
\drawline\fermion[\W\REG](3000,14000)[3000]
\drawline\fermion[\W\REG](3000,12000)[3000]

\drawline\fermion[\W\REG](3000,8000)[3000]
\drawline\fermion[\W\REG](3000,6000)[3000]

\drawline\fermion[\E\REG](8000,14000)[3000]
\drawline\fermion[\E\REG](8000,12000)[3000]

\drawline\fermion[\E\REG](8000,8000)[3000]
\drawline\fermion[\E\REG](8000,6000)[3000]

\put(13000,10000){$ = $}
\drawline\fermion[\E\REG](15000,14000)[6000]
\drawline\fermion[\E\REG](15000,12000)[6000]

\drawline\fermion[\E\REG](15000,8000)[6000]
\drawline\fermion[\E\REG](15000,6000)[6000]
\put(23000,10000){$ + $}
\drawline\fermion[\E\REG](32000,5000)[5000]
\drawline\fermion[\N\REG](\particlebackx,\particlebacky)[10000]
\drawline\fermion[\W\REG](\particlebackx,\particlebacky)[5000]
\drawline\fermion[\S\REG](\particlebackx,\particlebacky)[10000]
\put(34000,10000){$ f $}
\drawline\fermion[\W\REG](32000,14000)[5000]
\drawline\fermion[\W\REG](32000,12000)[5000]

\drawline\fermion[\W\REG](32000,8000)[5000]
\drawline\fermion[\W\REG](32000,6000)[5000]

\drawline\fermion[\E\REG](37000,14000)[3000]
\drawline\fermion[\E\REG](37000,12000)[3000]

\drawline\fermion[\E\REG](37000,8000)[3000]
\drawline\fermion[\E\REG](37000,6000)[3000]

\drawline\photon[\N\REG](28000,8000)[4]
\put(26000,14000){$i_r$}
\put(41000,14000){$\overline i_r$}

\put(26000,12000){$j$}
\put(41000,12000){$\overline j$}

\put(26000,8000){$i$}
\put(41000,8000){$\overline i$}

\put(26000,6000){$i_1$}
\put(41000,6000){$\overline i_1$}

\put(1500 ,13500){.}
\put(1500 ,13000){.}
\put(1500 , 12500){.}
\put(1500 ,10500){.}
\put(1500 ,10000){.}
\put(1500 , 9500){.}
\put(1500 ,7500){.}
\put(1500 ,7000){.}
\put(1500 ,6500){.}

\put(9500 ,13500){.}
\put(9500 ,13000){.}
\put(9500 , 12500){.}
\put(9500 ,10500){.}
\put(9500 ,10000){.}
\put(9500 , 9500){.}
\put(9500 ,7500){.}
\put(9500 ,7000){.}
\put(9500 ,6500){.}

\put(18000 ,13500){.}
\put(18000 ,13000){.}
\put(18000 , 12500){.}
\put(18000 ,10500){.}
\put(18000 ,10000){.}
\put(18000 , 9500){.}
\put(18000 ,7500){.}
\put(18000 ,7000){.}
\put(18000 ,6500){.}

\put(30500 ,13500){.}
\put(30500 ,13000){.}
\put(30500 , 12500){.}
\put(30500 ,10500){.}
\put(30500 ,10000){.}
\put(30500 , 9500){.}
\put(30500 ,7500){.}
\put(30500 ,7000){.}
\put(30500 ,6500){.}

\put(38500 ,13500){.}
\put(38500 ,13000){.}
\put(38500 , 12500){.}

\put(38500 ,10500){.}
\put(38500 ,10000){.}
\put(38500 , 9500){.}
\put(38500 ,7500){.}
\put(38500 ,7000){.}
\put(38500 ,6500){.}

\put(40000,0){\sl Fig. 1}
\end{picture}
\vspace*{2cm}

\begin{picture}(30000,20000)
\drawline\fermion[\E\REG](3000,10000)[6000]
\drawline\photon[\N\REG](6000,10000)[3]
\put(4000,8000){$i\  \kappa $}

\put(8000,8000){$ i^{\prime } \ \kappa^{\prime } $}
\put(6000,4000){\sl a}
\drawline\fermion[\E\REG](14000, 10000)[12000]
\drawline\photon[\N\REG](20000,10000)[3]
\drawline\fermion[\E\REG](14000, 13000)[12000]
\put(15000,8000){$i\    \kappa_i $}
\put(15000,14000){$j \  \kappa_j $}
\put(25000,8000){$i^{\prime }  \  \kappa_i^{\prime } $}
\put(25000,14000){$j^{\prime } \  \kappa_j^{\prime } $}
\put(20000,4000){\sl b}

\put(30000,0){\sl Fig. 2}

\end{picture}

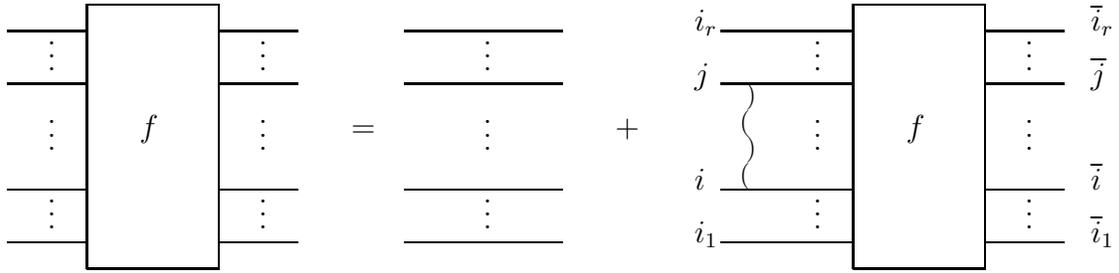
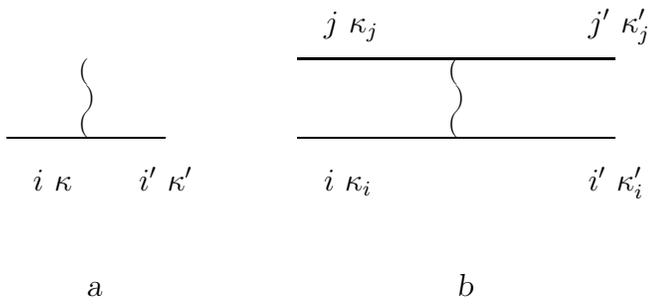
\end{document}